\documentclass[aps, pre, superscriptaddress, longbibliography, reprint]{revtex4-2}

\usepackage{graphicx}
\usepackage{tabularx}
\usepackage{adjustbox}
\usepackage{multirow}
\usepackage{booktabs}
\usepackage{array}
\usepackage{bm}
\usepackage{dcolumn}
\usepackage{amssymb,amsmath}
\usepackage{color}
\usepackage{siunitx}
\usepackage{enumitem}
\usepackage{longtable}
\usepackage[utf8]{inputenc}
\usepackage[T1]{fontenc}
\usepackage{textcomp}
\usepackage{amsbsy}
\usepackage{amsfonts}
\usepackage{dsfont}
\usepackage{mathtools}
\usepackage{braket} 
\usepackage{gensymb}
\usepackage{placeins}
\usepackage{dashrule}
\usepackage{xcolor}
\usepackage{float}
\usepackage{ulem}
\usepackage{subfigure}

\usepackage[colorlinks = true, allcolors = blue]{hyperref}

\usepackage{lipsum}
\newcommand{\cOut}[1]{}
\newcommand{\figref}[2]{\hyperref[#1]{\ref{#1}(#2)}}

\begin{document}


\title{All-electrical operation of a Curie-switch at room temperature}

\author{Vadym Iurchuk}
\email[e-mail: ]{v.iurchuk@hzdr.de}
\affiliation{Institute of Ion Beam Physics and Materials Research, Helmholtz-Zentrum Dresden-Rossendorf, 01328 Dresden, Germany}

\author{Oleksii Kozlov}
\affiliation{Institute of Magnetism, NASU and MESU, 03142 Kyiv, Ukraine}
\affiliation{National Technical University of Ukraine “Igor Sikorsky Kyiv Polytechnic Institute”, Prospekt Peremohy 37, 03056 Kyiv, Ukraine}

\author{Serhii Sorokin}
\affiliation{Institute of Ion Beam Physics and Materials Research, Helmholtz-Zentrum Dresden-Rossendorf, 01328 Dresden, Germany}

\author{Shengqiang Zhou}
\affiliation{Institute of Ion Beam Physics and Materials Research, Helmholtz-Zentrum Dresden-Rossendorf, 01328 Dresden, Germany}

\author{J\"urgen Lindner}
\affiliation{Institute of Ion Beam Physics and Materials Research, Helmholtz-Zentrum Dresden-Rossendorf, 01328 Dresden, Germany}

\author{Serhii Reshetniak}
\affiliation{Institute of Magnetism, NASU and MESU, 03142 Kyiv, Ukraine}
\affiliation{National Technical University of Ukraine “Igor Sikorsky Kyiv Polytechnic Institute”, Prospekt Peremohy 37, 03056 Kyiv, Ukraine}

\author{Anatolii Kravets}
\affiliation{Institute of Magnetism, NASU and MESU, 03142 Kyiv, Ukraine}
\affiliation{Nanostructure Physics, Royal Institute of Technology, 10691, Stockholm, Sweden}

\author{Dmytro Polishchuk}
\affiliation{Institute of Magnetism, NASU and MESU, 03142 Kyiv, Ukraine}
\affiliation{Nanostructure Physics, Royal Institute of Technology, 10691, Stockholm, Sweden}

\author{Vladislav Korenivski}
\affiliation{Nanostructure Physics, Royal Institute of Technology, 10691, Stockholm, Sweden}

\date{\today}

\begin{abstract}
We present all-electrical operation of a Fe$_x$Cr$_{1-x}$-based Curie switch at room temperature. More specifically, we study the current-induced thermally-driven transition from ferromagnetic to antiferromagnetic Ruderman-Kittel-Kasuya-Yosida (RKKY) indirect coupling in a Fe/Cr/Fe$_{17.5}$Cr$_{82.5}$/Cr/Fe multilayer. Magnetometry measurements at different temperatures show that the transition from the ferromagnetic to the antiferromagnetic coupling at zero field is observed at $\sim$325K. Analytical modelling confirms that the observed temperature-dependent transition from indirect ferromagnetic to indirect antiferromangetic interlayer exchange coupling originates from the modification of the effective interlayer exchange constant through the ferromagnetic-to-paramagnetic transition in the Fe$_{17.5}$Cr$_{82.5}$ spacer with minor contributions from the thermally-driven variations of the magnetization and magnetic anisotropy of the Fe layers. Room-temperature current-in-plane magnetotransport measurements on the patterned Fe/Cr/Fe$_{17.5}$Cr$_{82.5}$/Cr/Fe strips show the transition from the 'low-resistance' parallel to the 'high-resistance' antiparallel remanent magnetization configuration, upon increased probing current density. Quantitative comparison of the switching fields, obtained by magnetometry and magnetotransport, confirms that the Joule heating is the main mechanism responsible for the observed current-induced resistive switching.
\end{abstract}

\maketitle

\section{Introduction} 
Rapidly growing miniaturization demands in spintronics rely on the design and implementation of composite magnetic components, sensitive to various external control parameters (e.g. electrical current, magnetic field, temperature, pressure, etc)~\cite{dieny_opportunities_2020, bhatti_spintronics_2017, duine_synthetic_2018}. These components are considered as building blocks of emerging multifunctional spintronic devices with magnetoresistive readout, i.e magnetic sensors, magnetoresistive memories (MRAM) and spin-torque nano-oscillators (STNO)~\cite{chavent_multifunctional_2020,sousa_2020,ma_2021}. 

Retrospectively, \textit{interlayer exchange coupling} (IEC) was the basic phenomenon leading to the discovery of giant magnetoresistance (GMR)~\cite{Grunberg_GMR_1989, Fert_GMR_1988} and its subsequent industrial adoption in hard-drive read-heads, that kick-started the field of spintronics. This effect allows coupling between the constituent magnetic layers in magnetic multilayer stacks via conduction electrons of the non-magnetic metallic spacers~\cite{Stiles_interlayer_2004} through the so-called Ruderman–Kittel–Kasuya–Yosida (RKKY) interaction \cite{ruderman_kittel_RKKY_1954, kasuya_RKKY_1956, yosida_RKKY_1957}. The sign and the magnitude of this interaction, namely, whether ferromagnetic or antiferromagnetic arrangement of the magnetic layers is favored, and the strength of the coupling, depend on various intrinsic parameters: the constituent materials, magnetic and spacer layers thicknesses, the quality of the interface between the layers, etc~\cite{bruno_theory_1995}. Naturally, once the multilayer sample is fabricated, on-demand tuning of the RKKY coupling, which is desirable to cover diverse application directions of coupled magnetic multilayers, is not straightforward. Varying the temperature of the multilayer allows for a manipulation of the IEC strength, being presumably the only reasonable extrinsic parameter for tuning the RKKY interaction~\cite{bruno_theory_1995}. One of the alternative ways is using a specific spacer material with the Curie temperature in a vicinity of the room temperature. In such spacers, the thermally-induced transition from ferromagnetic to paramagnetic state alters the effective spacer thickness leading to a modification of the IEC.

Fe$_x$Cr$_{1-x}$ binary alloys are widely known diluted ferromagnets, which exhibit a temperature-dependent magnetic order-disorder phase transition with the Curie temperatures dependent on the alloy composition~\cite{babicIronMomentChromiumrich1980,burkeSuperparamagnetismCharacterMagnetic1978}. Recently, these compounds were successfully integrated into the so-called Curie switches -- the magnetic multilayers with thermally controlled IEC. More specifically, in Fe/Fe$_x$Cr$_{1-x}$/Fe-based multilayers, the transition from ferromagnetic-like to antiferromagnetic-like RKKY coupling through the Fe$_x$Cr$_{1-x}$ spacer can be finely controlled by varying either the temperature or the composition of the Fe$_x$Cr$_{1-x}$ spacer~\cite{polishchuk_thermally_2017}. Moreover using composite Cr/Fe$_x$Cr$_{1-x}$/Cr spacers with optimized thicknesses of the Cr layers was shown to sufficiently improve the thermally controlled switching of indirect IEC~\cite{polishchuk_thermally_2017} enabling its potential application in magnetic refrigeration~\cite{polishchuk_giant_2018} or thermal gating of spin waves~\cite{polishchuk_thermal_2021}. 

Here, we present a study of the thermally-controlled switching of the indirect IEC in Fe/Cr/Fe$_{17.5}$Cr$_{82.5}$/Cr/Fe-based Curie switch. First, we use temperature-dependent vibrating sample magnetometry (VSM) measurements to demonstrate the transition from the ferromagnetic-like to the antiferromagnetic-like RKKY coupling at remanence upon crossing the Curie temperature of the Fe$_{17.5}$Cr$_{82.5}$ spacer. Analytical modelling, based on the magnetic energy minimization, confirms that the origin of the observed transition is the temperature dependence of the IEC constant with minor contributions from thermally-related variations of the magnetization and magnetic anisotropy. For given composition of the Fe$_{17.5}$Cr$_{82.5}$ spacer, the transition temperature was estimated to be $\sim$325 K, allowing for the current-driven switching of the IEC at room temperature via Joule heating. Finally, current-in-plane magnetotransport measurements on the patterned Fe/Cr/Fe$_{17.5}$Cr$_{82.5}$/Cr/Fe spin valves confirm the current-induced reversible control of the IEC at remanence and demonstrate the magnetoresistive switching between the 'low-resistance' parallel and 'high resistance' antiparallel states at zero-field. 

\begin{figure}[t]
\centering
    \includegraphics[width=0.5\textwidth]{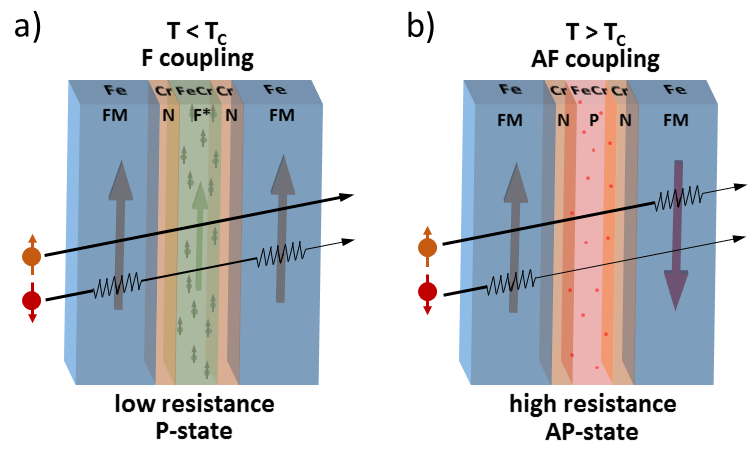}
    \caption{Schematics of the current-perpendicular-to-plane spin-dependent electron transport through the FM/N/F*/N/FM multilayer with thermally controlled indirect IEC through the composite N/F*/N spacer. Here, FM = Fe, F* =  Fe$_{x}$Cr$_{1-x}$ and N = Cr. Red arrows denote the magnetization directions in the FM layers. Green arrow in (a) shows the magnetization of the F* layer. Black arrows show the flow of the spin-polarized electrons through the multilayer. (a) Below the Curie temperature the spacer F* is ferromagnetic, therefore imposing a ferromagnetic-like RKKY coupling between the FM layers. Reduced scattering of the electrons with the spin parallel to the FM magnetization results in the 'low resistance' state of the multilayer. (b) Above the Curie temperature the spacer is paramagnetic (P), which leads to the antiferromagnetic-like indirect IEC. The multilayer is therefore in the 'high resistance' state due to the enhanced scattering of the electrons in both spin-dependent channels.}
    \label{Fig1}
\end{figure}

\section{Concept of thermally induced resistance switching}
Fig.~\ref{Fig1} shows the schematics of the spin-dependent scattering of the electrons flowing through the Fe/Cr/Fe$_{x}$Cr$_{1-x}$/Cr/Fe multilayer below the Curie temperature ($T<T_C$,  Fig.~\ref{Fig1}(a)) and above the Curie temperature ($T>T_C$, Fig.~\ref{Fig1}(b)). As described in details in~\cite{polishchuk_thermally_2017}, in such structures, the IEC is temperature-dependent, since it is defined by the magnetic state of the Fe$_{x}$Cr$_{1-x}$ spacer. More specifically, for $T<T_C$, the Fe$_{x}$Cr$_{1-x}$ is ferromagnetically ordered allowing for the indirect ferromagnetic coupling to the adjacent Fe layers. This enforces \textit{parallel} (P) orientation of all magnetic layers of the trilayer. For this configuration, according to the resistor model of the magnetotransport through the spin valve~\cite{valet_1993}, a \textit{low resistance} state of the structure is expected due to the reduced scattering of the electrons with the spin polarization parallel to the magnetic moments of the constituent magnetic layers (see Fig.~\ref{Fig1}(a)).
Upon increasing temperature, the Fe$_{x}$Cr$_{1-x}$ spacer undergoes the second order magnetic phase transition in the vicinity of $T_C$, and becomes paramagnetic for $T>T_C$. At zero magnetic field, an antiferromagnetic-like indirect RKKY coupling between the Fe layers is therefore expected. Thus, the \textit{antiparralel} (AP) orientation of the Fe layers magnetic moments leads to the \textit{high resistance} state, as now the electrons with both spin polarization directions scatter significantly when flowing through the spin valve. One has to note that the described concept is equally valid for both current-perpendicular-to-plane and current-in-plane transport geometries.

This approach allows for a magnetic field-free thermally-controlled magnetotransport through the Fe$_{x}$Cr$_{1-x}$-based multilayers with the possibility to address the resistance state of the structure by controlling its interlayer magnetic coupling. In this work, we propose to take advantage of current-induced Joule heating as the effective extrinsic means of controlling the temperature of the sample.

\begin{figure*}[t]
    \includegraphics[width=0.75\textwidth]{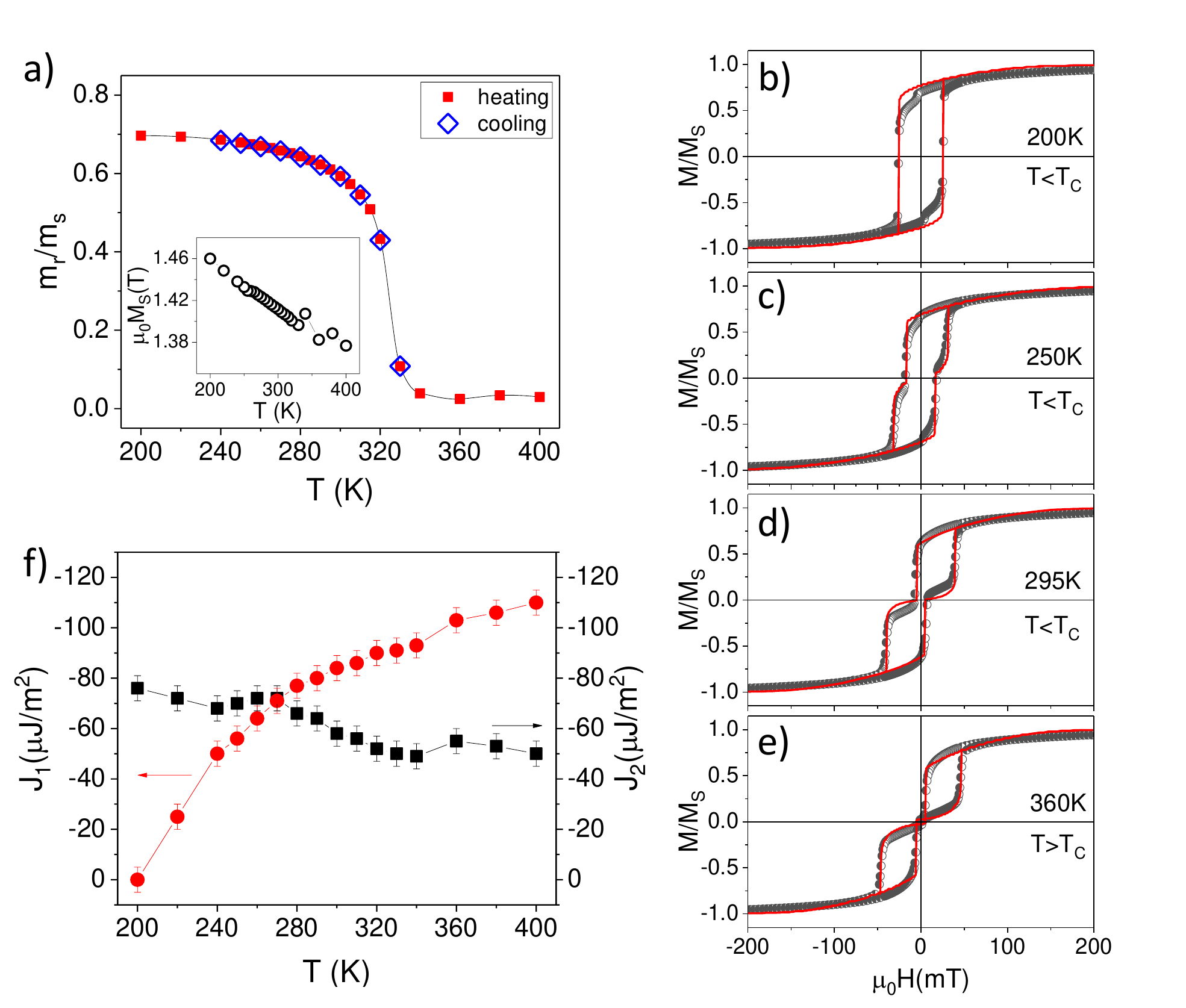}
    \caption{(a) Normalized remanent magnetic moment of the Fe/Cr/Fe$_{17.5}$Cr$_{82.5}$/Cr/Fe multilayer as a function of increasing (red squares) and decreasing (blue circles) temperature. The inset shows the saturation magnetization $M_s$ vs. temperature. (b--e) Grey circles: magnetic hysteresis loops measured at 200 K (b), 250 K (c), 295~K (d) and 360 K (e). Red lines: Analytical fits to the experimental data for given sets of the temperature-dependent interlayer coupling constants (see section~\ref{Model} for details).
    The shape of the magnetic hysteresis and the magnetic moment value at remanence clearly indicates ferromagnetic-like coupling below Curie temperature and antiferromagnetic-like coupling above Curie temperature at zero field.
    (f) Bilinear and biquadratic IEC constants $J_1$ and $J_2$ as a function of temperature extracted from the analytical hysteresis loops.}
    \label{Fig2}
\end{figure*}

\section{Samples and experimental details}
The Fe(2)/Cr(0.4)/Fe$_{17.5}$Cr$_{82.5}$(0.9)/Cr(0.4)/Fe(2) multilayers (thicknesses are given in nm) were grown on Ar$^+$ pre-etched undoped (100)Si substrates by dc magnetron sputtering (by AJA International). The Fe$_{17.5}$Cr$_{82.5}$ layers were deposited using co-sputtering from separate Fe and Cr targets.

The temperature-dependent vibrating sample magnetometry (VSM) measurements were performed in the Quantum Design Magnetic Properties Measurement System (MPMS3) using Superconducting Quantum Interference Device (SQUID). For all measurements, the magnetic field was applied in the sample plane.

For the magnetotransport measurements, the full-sheet films were patterned into 300 $\mu$m long strips with various widths (9, 7 and 4 $\mu$m), using conventional UV lithography followed by the reactive Ar$^+$ ion etching. To allow for the electrical access to the strips, Cr(5~nm)/Au(125~nm) contact pads were fabricated by a standard process including UV lithography, metal deposition by e-beam evaporation and lift-off.
Two-point magnetoresistance measurements were carried out in a standard magnetotransport setup by probing the voltage drop across the sample as a function of the magnetic field at constant dc current passed through the strip.

\section{Results and discussion}

\subsection{Temperature-dependent VSM measurements} \label{VSM}

To verify the temperature-induced transition from ferromagnetic-like to antiferromagnetic-like IEC at zero magnetic field, we have measured the in-plane magnetic hysteresis loops of the full-sheet Fe/Cr/Fe$_{17.5}$Cr$_{82.5}$/Cr/Fe multilayer at different temperatures ranging from 200 K to 400 K. 
Fig.~\ref{Fig2}(a) shows the remanent magnetic moment $m_r$ normalized by the saturated moment $m_s$ of the multilayer as a function of the temperature $T$. For each temperature point, the full hysteresis loop was measured starting from the saturation at high field ($\mu_0 H\sim$2 T), and the corresponding remanent magnetization $m_r$ is extracted at zero field after reducing the field from the positive saturation. Upon increasing temperature, a gradual decrease of the $m_r$ is observed first, followed by the drastic reduction to zero for $T >$ 325 K. The observed $m_r(T)$ dependence indicates a qualitative transition between different magnetic states of the multilayer. Notably, no temperature-dependent hysteresis was detected, i.e. the $m_r(T)$ curve measured when heating the sample [red squares in Fig.~\ref{Fig2}(a)] is equivalent to the one recorded upon cooling [blue diamonds in Fig.~\ref{Fig2}(b)].

To get more insight onto the magnetic configuration at various $T$ values, we examined the hysteresis loops measured at $T=$ 200; 250; 295 and 360~K. The corresponding loops are shown in Fig.~\ref{Fig2}(b--d). At $T$ = 200~K (see Fig.~\ref{Fig2}(b)), the magnetic hysteresis loop exhibits a shape, typical for the ferromagnetically coupled systems, with a distinct magnetization switching at $|\mu_0 H_c|\approx$ 25~mT. This coercive field is mainly related to the magnetocrystalline anisotropy of the Fe layers~\cite{polishchukMagneticHysteresisNanostructures2018} and a possible presence of the pinning sites at the interfaces due to the polycrystalline nature of the sputtered films. Upon increasing the temperature, the shape of the hysteresis is modified mainly in the vicinity of the switching field range, developing a well-defined "plateau" with the close-to-zero value of the magnetic moment. This plateau indicates the field range where the antiferromagnetic-like coupling between the Fe layers can be stabilized at a given temperature, overcoming the effects of anisotropy and pinning. The onset temperature, where the plateau is first detected, is $\sim$240 K. The width of the observed plateau gradually increases with increasing temperature (see the hysteresis loops in Fig.~\ref{Fig2}(c--e)). The observed qualitative modification of the hysteresis is attributed to the temperature-dependent ferromagnetic-to-paramagnetic transition in the Fe$_{17.5}$Cr$_{82.5}$ spacer, which effectively alters the IEC between Fe layers from the ferromagnetic-like to the antiferromagnetic-like. Therefore, the decrease of the remanent magnetic moment with increasing temperature [Fig.~\ref{Fig2}(a)] originates from the zero-moment plateau widening due to the dominating antiferromagnetic-like coupling in the trilayer above the Curie temperature $T_C$, which is estimated to be $\sim$325 K. For $T > T_C$, the aniferromagnetic interlayer coupling can be stabilized at zero field [see the hysteresis loop at 360K in Fig.~\ref{Fig2}(e)].

The VSM measurements show that the saturation magnetization of the multilayer changes scarsely in the given temperature range [see inset in Fig.~\ref{Fig2}(a)]. Thus, the observed transition from the ferromagnetic to the antiferromagnetic coupling can be mainly attributed to the thermal variation of the effective IEC due to the ferromagnetic-to-paramagnetic transition in the Fe$_{17.5}$Cr$_{82.5}$ spacer.

\subsection{Analytical modelling of the temperature-dependent IEC} \label{Model}
To compare the strength of the IEC for different temperatures, we conducted an analytical modelling of the temperature-dependent magnetic hysteresis loops using a macrospin model similar to previously presented in~\cite[Eq.~2--5]{Belmeguenai_model_2007} and~\cite[Eq.~1--3]{sorokin_magnetization_2020}. Qualitatively, the IEC between two magnetic layers separated by a spacer can be described using the following relation~\cite[pp.~99--100]{Stiles_interlayer_2004}:

\begin{equation}
    E_{RKKY} = - J_1 \frac{\textbf{M}_1 \cdot \textbf{M}_2}{M_{s1}M_{s2}} - J_2 \left( \frac{\textbf{M}_1 \cdot \textbf{M}_2}{M_{s1}M_{s2}} \right)^2 
    \label{eq:RKKY_qualitative}
\end{equation}

Here, $J_1$ and $J_2$ are the so-called bilinear and biquadratic coupling constants, $M_1$ and $M_2$ are the magnetizations of the constituent layers, and $M_{s1}$, $M_{s2}$ are the corresponding saturation magnetizations. The first term in the Eq.~\ref{eq:RKKY_qualitative} represents the indirect exchange interaction between the layers through the conduction electrons. The second term is included to account for the structural inhomogeneities of the real sample (such as interface roughness, etc) or the finite temperature, as often required to correctly describe real experimental results~\cite[p.~119]{Stiles_interlayer_2004}.

The total magnetic energy per unit area, including  Zeeman energy, uniaxial anisotropy energy, and bilinear and biquadratic RKKY coupling contributions, can be written as follows:

\begin{multline}
    \epsilon_{tot} = \sum_{i=1,2}d_i \left[ -M_{s_i} \mu_0 H \cos(\phi_i) - K_{u_i} \cos^2(\alpha_i - \phi_i) \right] \\ - J_1 \cos(\phi_1 - \phi_2) - J_2 \cos^2(\phi_1 - \phi_2),
    \label{eq:total_static_energy}
\end{multline}
\\
where $d_i$ are the thicknesses of the magnetic layers, $M_{s_i}$ are the corresponding saturation magnetizations, $\mu_0$ is the vacuum permeability, $H$ is the magnitude of the applied magnetic field, $\phi_i$ are the angles between the magnetizations and the applied field direction, $K_{u_i}$ are uniaxial magnetocrystalline anisotropy energies per unit volume and $\alpha_i$ are the angles between the anisotopy axes and the applied field direction. In our case both Fe layers have the same thickness, therefore we assume that $d_1 = d_2 = d$ and $M_{s1} = M_{s2} = M_s$.

Minimizing the total energy by solving $\frac{\partial \epsilon_{tot}}{\partial \phi_i } = 0$ for different values of the applied field $H$ allows for a determination of the equilibrium angles $\phi_{i_{eq}}$ for all possible magnetic configurations states from AP-coupled to saturation. Hence, the normalized magnetization $M(H)/M_s$ as a function of the applied field can be expressed as:

\begin{equation}
    \frac{M(H)}{M_s} = \frac{1}{2} \left( \cos\phi_{1_{eq}} + \cos\phi_{2_{eq}} \right),
\label{moment}
\end{equation}
\\
with $\phi_{i_{eq}} = f(H, J_1, J_2, K_{u_i}, \alpha_i)$ being the equilibrium angles for given values of applied magnetic field $H$ and fitting parameters $J_1$, $J_2$, $K_{u_i}$, $\alpha_i$.

We calculate the hysteresis loops by fitting the Eq.~\ref{moment} to the experimental data obtained by magnetometry. For each temperature value (except 360, 380 and 400~K), we use $K_{u_1}$ = $K_{u_2}$ = 28~kJ/m$^3$, $\alpha_1$ = $\pi$/5, $\alpha_2$ = --$\pi$/10, and the $M_s$ values, extracted from the VSM [see inset in Fig.~\ref{Fig2}(a)]. For $T$ = 360, 380 and 400~K, lower $K_u$ values were used (27, 26 and 24~kJ/m$^3$ respectively) to account for the temperature-induced softening of the magnetocrystalline anisotropy. In addition, to ensure better fit, small deviations ($\leq$~15 deg) of the angles $\alpha_i$ were introduced during the total energy minimization. To mimic the thermally induced modification of the IEC, different values of $J_1$ and $J_2$ were used to calculate the hysteresis at corresponding temperatures [red lines in Fig.~\ref{Fig2}(b--e)].
Fig.~\ref{Fig2}(f) shows the values of $J_1$ and $J_2$ extracted form the analytical modelling of the the experimental data by energy minimization for the given values of $M_s$, $K_{u_i}$ and $\alpha_i$. One can see that at $T$=200~K, $J_1 \approx$~0, which corresponds to the dominant ferromagnetic-like IEC in the trilayer due to the indirect coupling through the ferromagnetic Fe$_{17.5}$Cr$_{82.5}$ spacer. Upon increasing temperature, the absolute value of $J_1$ gradually increases reaching $J_1 \approx$--100~$\mu$J/m$^2$ for the temperatures above $T_C$, while $J_2$ exhibits only small decrease.
This behavior can be understood from the phenomenology of the coupling constants $J_1$ and $J_2$. The type and strength of the interlayer coupling is mainly defined by the sign and magnitude of the bilinear coupling constant $J_1$~\cite{Stiles_interlayer_2004}. Increased absolute values of $J_1$ at increased temperatures indicate the transition from the ferromagnetic-like IEC below $T_C$ (when the Fe$_{17.5}$Cr$_{82.5}$ spacer is still ferromagnetic) to the dominant indirect antiferromagnetic-like IEC through the paramagnetic Fe$_{17.5}$Cr$_{82.5}$ spacer above $T_C$.
On the other hand, the biquadratic coupling constant $J_2$ is phenomenologically introduced to account for the effects of the physical interface between the FM layers (e.g. roughness, grain distribution, etc)~\cite{Stiles_interlayer_2004}. Its weak dependence on the temperature suggests that the quality of the interface between the Fe layers is not significantly impacted by the phase transition in the Fe$_{17.5}$Cr$_{82.5}$ spacer due to the presence of the additional Cr layers at both Fe/Fe$_{17.5}$Cr$_{82.5}$ interfaces. One has to note, that relatively large values of $J_2$ suggest that the zero-field magnetic configuration below $T_C$ is not purely ferromagnetically aligned but rather canted, which is in argeement with relatively low values of the measured remanent magnetization ($\sim$~0.7) at low temperatures.

One has to comment that we were unable to obtain perfect fitting of the experimental hysteresis data with our model. The discrepancies mainly arise due to the model simplifications assuming only one interface between the FM layers, whereas in a real sample, and especially for $T \leqslant T_C$, two Fe/Fe$_{17.5}$Cr$_{82.5}$ interfaces have to be considered with two sets of IEC constants.
Nevertheless, a good agreement between the experimental and analytical data suggests that the temperature dependence of $J_1$ is the dominant factor contributing to the decrease of the remanent magnetic moment and, therefore, to the thermally-driven transition from the ferromagnetic-like to the antiferomagnetc-like IEC at zero-field.

We note that in addition to the pure RKKY-like IEC, a possible presence of the "loose spins"~\cite{slonczewskiOriginBiquadraticExchange1993} in the Fe$_{17.5}$Cr$_{82.5}$ spacer may contribute to the bilinear IEC constant $J_1$ and especially to the biquadratic constant $J_2$. Although our measurements do not allow to disentangle the two contributions, we argue that due to the presence of nonmagnetic pure-Cr layers at the Fe/Fe$_{17.5}$Cr$_{82.5}$ interface, the main path for IEC is the indirect RKKY via conduction electron-mediated exchange. Second order effects, including "loose spins", are possible in paramagnetic Fe$_{17.5}$Cr$_{82.5}$, which may alter the RKKY coupling strength in the AP state. Therefore, above $T_C$, the RKKY coupling may be weakened by the spin-flip scattering of the spin-polarized conduction electrons (carrying RKKY via the spacer) on the "paramagnetic Fe impurities" in Cr, which are interpreted as "loose spins". However, a weak $J_2$ dependence on the temperature suggests a minor contribution of the "loose spins" to the IEC.

\begin{figure}[h]
    \includegraphics[width=0.45\textwidth]{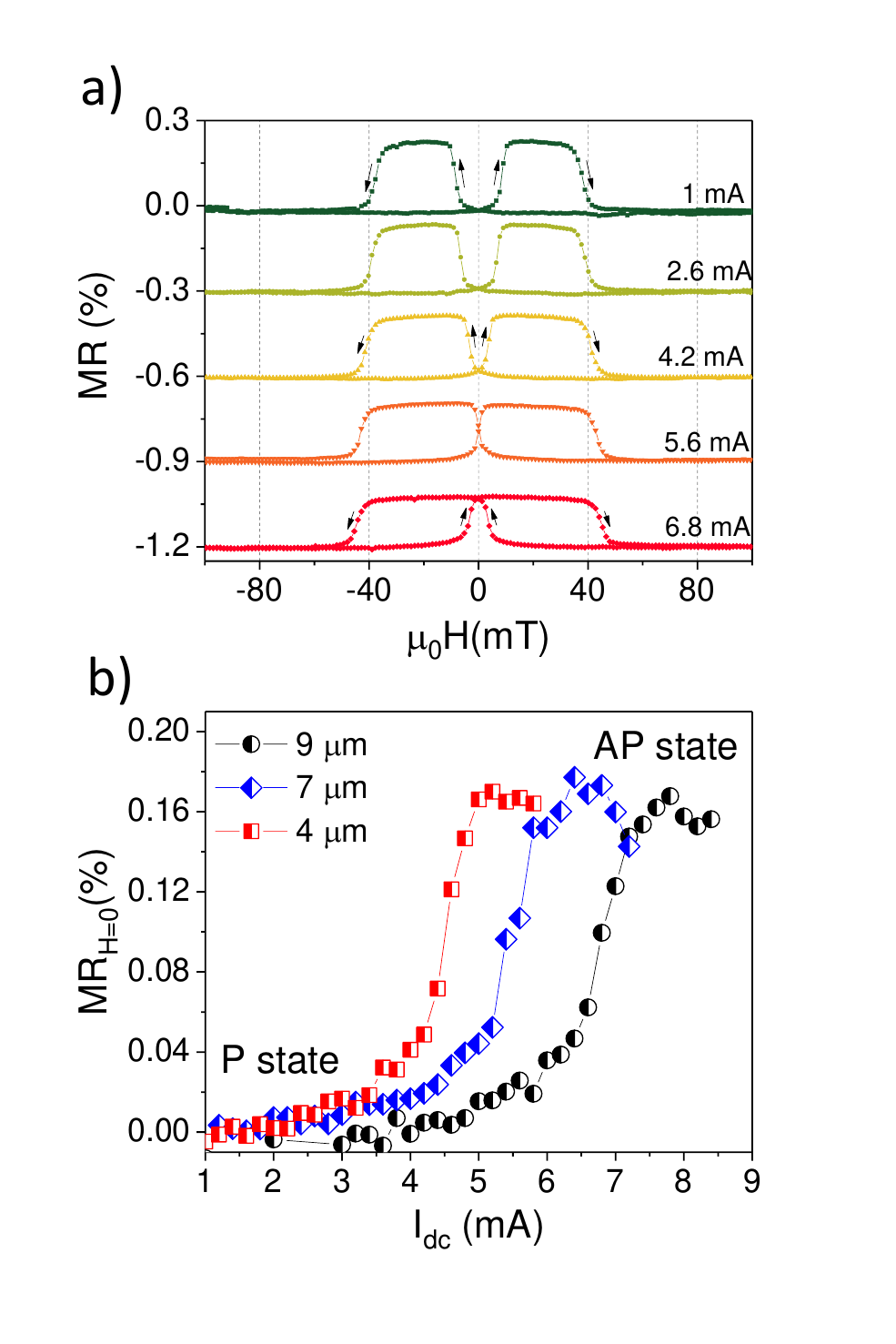}
    \caption{(a) Typical magnetoresistance loops measured on the 7-$\mu$m wide strip for $H \parallel I_{dc}$ and for the different values of the bias current injected into the device. The loops are offset by 0.4\%. Arrows serve to guide the resistance change  during the field sweeps. (b) Zero-field magnetoresistance values vs. dc current $I_{dc}$ extracted from the $MR(H)$ loops measured on the strips with different widths.}
    \label{Fig3}
\end{figure}

\subsection{Magnetotransport measurements} \label{MR}
Fig.~\ref{Fig3}(a) shows the magnetoresistance (MR) loops measured at room temperature ($\sim$294 K) on a 300~$\mu$m-long and 7~$\mu$m-wide strip for the magnetic field $H$ parallel to the probing current $I_{dc}$ applied along the bar length. For this geometry, the measured resistance change is mainly attributed to the GMR effect in the RKKY-coupled Fe/Cr/Fe$_{17.5}$Cr$_{82.5}$/Cr/Fe spin valve with minor contribution from the anisotropic magnetoresistance of Fe.

For relatively low $I_{dc}$ = 1 mA (top green loop), the MR dependence vs. applied field follows the magnetic hysteresis of Fig.~\ref{Fig2}(d) measured at 295~K. The shape of the MR loop reveals two distinct plateaus where the MR is maximum. These plateaus correspond to the field regions of the close-to-zero magnetic moment, where the antiferromagnetic-like RKKY coupling dominates. The corresponding P-to-AP and AP-to-P switching fields are $\pm$10 and $\pm$40 mT respectively, in good agreement with the magnetometry data of Figure~\ref{Fig2}(c).

Upon increasing bias current injected through the trilayer, the width of the high-resistance plateau increases as a result of the shift of the switching fields of the P-to-AP and AP-to-P transitions. More specifically, when the field is reduced from the saturation, we observe the reduction of the P-to-AP switching field and the corresponding AP-to-P switching field growth, for increased values of the dc current (see the MR loops for the corresponding $I_{dc}$ values in Fig.~\ref{Fig3}(a)). Eventually, the P-to-AP switching field approaches zero for the critical dc current $I_C$ = 5.4~mA, and becomes positive for the above-critical currents. This leads to the gradual increase of the zero-field MR as indicated by the MR loops measured for $I_{dc}$ = 4.2, 5.6 and 6.8 mA. The observed current-induced transition is attributed to the Joule heating of the Fe/Cr/Fe$_{17.5}$Cr$_{82.5}$/Cr/Fe spin valve acting similar to the conventional heating effects described in section~\ref{VSM}.

\begin{figure}[h]
    \includegraphics[width=0.45\textwidth]{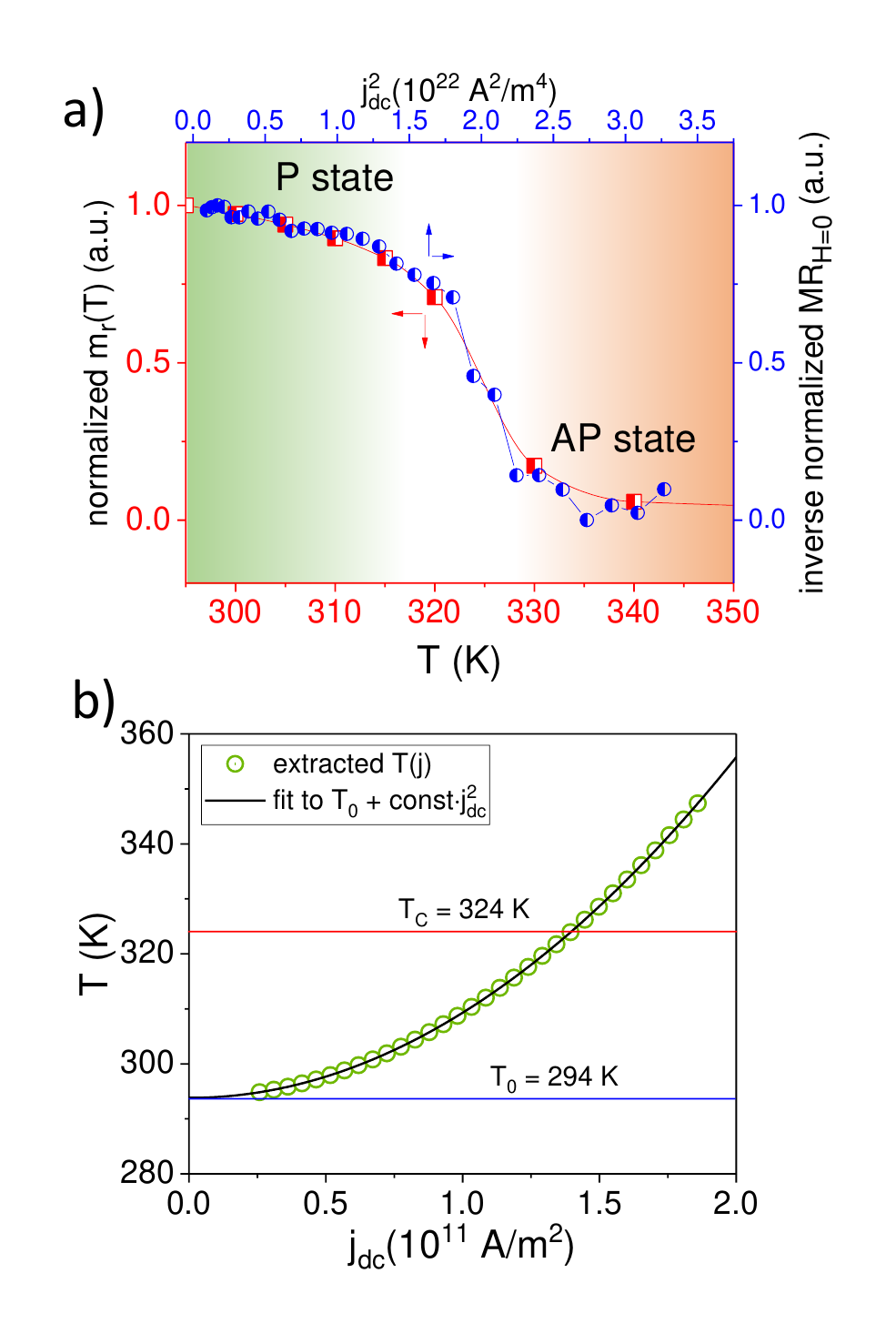}
    \caption{(a) Remanent magnetic moment vs. temperature (red circles) superimposed onto the inverted zero-field magnetoresistance vs. current density (blue squares) for 7-$\mu$m wide strip. Both quantities are scaled to [0;1] range for better visualization. (b) Temperature vs. current density dependence (green circles) extracted from the data of (a). Black solid line is the fit to the parabolic $T_0 + const \cdot j_{dc}^2$ equation, where $T_0$ is room temperature. The horizontal blue and red lines mark the room temperature and the Curie temperature of teh Fe$_{17.5}$Cr$_{82.5}$ spacer respectively.}
    \label{Fig4}
\end{figure} 

Fig.~\ref{Fig3}(b)) shows the zero-field magnetoresistance value extracted from the magnetotransport loops at different values of the dc current and for different widths of the spin valves. Notably, the critical current increase for the increased strip width is attributed to the geometric resistance effects, thus confirming that the origin of the P-to-AP transition is the current-induced Joule heating. Fig.~\ref{Fig4}(a) shows the normalized remanent magnetic moment vs. temperature (red circles) superposed onto the inverse normalized zero-field magnetoresistance of the 7-$\mu$m strip vs. squared current density (blue squares) taken from Fig.~\ref{Fig3}(b). An excellent correlation of these dependencies suggests that the current-driven transition from 'low resistance' P state to the 'high-resistance' AP state at room temperature originates from the transition from the ferromagnetic-like to the antiferromagnetic-like RKKY coupling between the Fe layers occurring for high currents and, hence, for the increased temperature of the spin valve. Fitting both dependencies by a sigmoidal function allows one to estimate the current-induced Joule heating of the spin valve device. The corresponding $T(j_{dc})$ dependence extracted from the data of Fig.~\ref{Fig4}(a) is shown in Fig.~\ref{Fig4}(b) as green circles, and fitted by a parabolic relation $T = T_0 + const \cdot j_{dc}^2$ (black line), where $T_0$ = 294~K is room temperature. From Fig.~\ref{Fig4}(a) and (b), one can estimate the critical current density $j_c$ = 1.39$\times$10$^{11}$~A/m$^2$, which corresponds to the Joule heating of the sample to the Curie temperature of $\sim$324~K.

These results enable a precise control of the magnetic state of such spin valve at room temperature and zero magnetic field by the Joule heating allowing for the resistive P-to-AP switching by solely injecting moderate dc currents through the spin valve.
One has to note that the fine engineering of the Fe$_{x}$Cr$_{1-x}$ material parameters, i.e. optimization of the switching fields and the Curie temperature, allows for an on-demand precise tuning of the operation temperature ranges and, therefore, the current densities needed for the resistive switching of the designed Fe/Cr/Fe$_{x}$Cr$_{1-x}$/Cr/Fe spin valves.

One the other hand, the scaling of the Curie-switch-based devices down to the sub-micron size allows not only for the reduction of the switching current density but is expected to considerably decrease the switching time. For example, the characteristic time needed to heat a thermally-assisted MRAM cell above the blocking temperature is few ns~\cite{prejbeanuThermallyAssistedMRAM2007}, whereas typical times for the nanoscale phase change memory devices (also based on heating/cooling processes) are shown to be below 10~ns~\cite{legalloEvidenceThermallyAssisted2016}.

From the application point of view, finely controllable IEC in magnetic multilayers brings benefits for emerging spintronic devices due to the potential facilitation of the magnetization switching process in magnetoresistive memory cells~\cite{strelkov_2018, chaventSteadyStateDynamics2016}. Indeed, such a Curie-switch embedded into MRAM nanopillar allows for electrically controlled change of the fringing field from maximum (P state) to close-to-zero (AP state), which may potentially act as a ‘write head’ for toggling a free layer in MRAM and other devices where modulation of a local field is needed.

Another interesting application of Curie-switches is a thermal control of the ferromagnetic and antiferromagnetic magnon modes, which mediate the coupling between the ferromagnetic layers through the phase transition in an antiferromagnetic (AF) spacer of F/AF/F trilayers~\cite{polishchuk_thermal_2021}. Recently, a current-induced resistive switching in F/AF/F trilayers was shown~\cite{wuCurrentinducedNeelOrder2022} with the switching current densities comparable to the observed in the present study ($\sim$10$^{11}$~A/m$^2$), making both approaches suitable for further optimization.

Finally, the proposed electrically tunable Curie switches may be employed as composite free layers in spin-torque oscillators introducing an extra room for dynamical tuning. In such STNO, the dc current is expected to not only shift the rf generation frequency due to the non-isochronous property of a STNO, but also change qualitatively the mode character (from "optic-like" to "acoustic-like" accompanied by the corresponding frequency jump) due to the modified coupling between the constituent magnetic layers. Therefore, such electrically controlled interlayer coupling through the phase transition in the Fe$_{17.5}$Cr$_{82.5}$ spacer at room-temperature allows for the dual-band STNO based on a single magnetic stack with current-induced switching between the two frequency bands.

\section{Conclusions}
We demonstrated all-electrical operation of the Fe$_{x}$Cr$_{1-x}$-based Curie-switch at room temperature. More specifically, we showed a current-induced thermally-driven transition from ferromagnetic to antiferromagnetic RKKY coupling in Fe/Cr/Fe$_{17.5}$Cr$_{82.5}$/Cr/Fe multilayers. Using temperature-dependent magnetometry measurements, we determined that the transition from the ferromagnetic to the antiferromagnetic coupling at zero field occurs at $\sim$325K. Subsequently, we showed that the thermally-driven effects may be stimulated by means of Joule heating upon injecting a dc current into the Fe/Cr/Fe$_{17.5}$Cr$_{82.5}$/Cr/Fe microstrip. Magnetotransport measurements confirm the current-induced reversible control of the IEC and demonstrate the magnetoresistive switching between the 'low-resistance' P and 'high resistance' AP states at zero-field. 
As suggested by the analytical calculations, based on the energy minimization, the observed transition mainly originates from the temperature dependence of the exchange constant $J_1$. 
Fine control over the IEC in magnetic multilayers is expected to facilitate the magnetization switching process in magnetoresistive memory cells, as well as introduce an extra room for thermally-controlled dynamical tuning of spin-torque nano-oscillators.

\begin{acknowledgments}
Support from the Nanofabrication Facilities Rossendorf (NanoFaRo) at the IBC is gratefully acknowledged. The authors acknowledge support from the National Academy of Sciences of Ukraine (projects 0118U003265 and 0120U100457), the Swedish Research Council (VR 2018-03526), the Olle Engkvist Foundation (project 2020-207-0460), the Volkswagen Foundation (Grant No. 97758), the Wenner-Gren Foundation (grant GFU2022-0011), and the Swedish Strategic Research Council (SSF UKR22-0050).
\end{acknowledgments}

\bibliography{References}

\begin{thebibliography}{28}%
\makeatletter
\providecommand \@ifxundefined [1]{%
 \@ifx{#1\undefined}
}%
\providecommand \@ifnum [1]{%
 \ifnum #1\expandafter \@firstoftwo
 \else \expandafter \@secondoftwo
 \fi
}%
\providecommand \@ifx [1]{%
 \ifx #1\expandafter \@firstoftwo
 \else \expandafter \@secondoftwo
 \fi
}%
\providecommand \natexlab [1]{#1}%
\providecommand \enquote  [1]{``#1''}%
\providecommand \bibnamefont  [1]{#1}%
\providecommand \bibfnamefont [1]{#1}%
\providecommand \citenamefont [1]{#1}%
\providecommand \href@noop [0]{\@secondoftwo}%
\providecommand \href [0]{\begingroup \@sanitize@url \@href}%
\providecommand \@href[1]{\@@startlink{#1}\@@href}%
\providecommand \@@href[1]{\endgroup#1\@@endlink}%
\providecommand \@sanitize@url [0]{\catcode `\\12\catcode `\$12\catcode
  `\&12\catcode `\#12\catcode `\^12\catcode `\_12\catcode `\%12\relax}%
\providecommand \@@startlink[1]{}%
\providecommand \@@endlink[0]{}%
\providecommand \url  [0]{\begingroup\@sanitize@url \@url }%
\providecommand \@url [1]{\endgroup\@href {#1}{\urlprefix }}%
\providecommand \urlprefix  [0]{URL }%
\providecommand \Eprint [0]{\href }%
\providecommand \doibase [0]{https://doi.org/}%
\providecommand \selectlanguage [0]{\@gobble}%
\providecommand \bibinfo  [0]{\@secondoftwo}%
\providecommand \bibfield  [0]{\@secondoftwo}%
\providecommand \translation [1]{[#1]}%
\providecommand \BibitemOpen [0]{}%
\providecommand \bibitemStop [0]{}%
\providecommand \bibitemNoStop [0]{.\EOS\space}%
\providecommand \EOS [0]{\spacefactor3000\relax}%
\providecommand \BibitemShut  [1]{\csname bibitem#1\endcsname}%
\let\auto@bib@innerbib\@empty
\bibitem [{\citenamefont {Dieny}\ \emph {et~al.}(2020)\citenamefont {Dieny},
  \citenamefont {Prejbeanu}, \citenamefont {Garello}, \citenamefont
  {Gambardella}, \citenamefont {Freitas}, \citenamefont {Lehndorff},
  \citenamefont {Raberg}, \citenamefont {Ebels}, \citenamefont {Demokritov},
  \citenamefont {Akerman}, \citenamefont {Deac}, \citenamefont {Pirro},
  \citenamefont {Adelmann}, \citenamefont {Anane}, \citenamefont {Chumak},
  \citenamefont {Hirohata}, \citenamefont {Mangin}, \citenamefont {Valenzuela},
  \citenamefont {Onbaşlı}, \citenamefont {d’Aquino}, \citenamefont
  {Prenat}, \citenamefont {Finocchio}, \citenamefont {Lopez-Diaz},
  \citenamefont {Chantrell}, \citenamefont {Chubykalo-Fesenko},\ and\
  \citenamefont {Bortolotti}}]{dieny_opportunities_2020}%
  \BibitemOpen
  \bibfield  {author} {\bibinfo {author} {\bibfnamefont {B.}~\bibnamefont
  {Dieny}}, \bibinfo {author} {\bibfnamefont {I.~L.}\ \bibnamefont
  {Prejbeanu}}, \bibinfo {author} {\bibfnamefont {K.}~\bibnamefont {Garello}},
  \bibinfo {author} {\bibfnamefont {P.}~\bibnamefont {Gambardella}}, \bibinfo
  {author} {\bibfnamefont {P.}~\bibnamefont {Freitas}}, \bibinfo {author}
  {\bibfnamefont {R.}~\bibnamefont {Lehndorff}}, \bibinfo {author}
  {\bibfnamefont {W.}~\bibnamefont {Raberg}}, \bibinfo {author} {\bibfnamefont
  {U.}~\bibnamefont {Ebels}}, \bibinfo {author} {\bibfnamefont {S.~O.}\
  \bibnamefont {Demokritov}}, \bibinfo {author} {\bibfnamefont
  {J.}~\bibnamefont {Akerman}}, \bibinfo {author} {\bibfnamefont
  {A.}~\bibnamefont {Deac}}, \bibinfo {author} {\bibfnamefont {P.}~\bibnamefont
  {Pirro}}, \bibinfo {author} {\bibfnamefont {C.}~\bibnamefont {Adelmann}},
  \bibinfo {author} {\bibfnamefont {A.}~\bibnamefont {Anane}}, \bibinfo
  {author} {\bibfnamefont {A.~V.}\ \bibnamefont {Chumak}}, \bibinfo {author}
  {\bibfnamefont {A.}~\bibnamefont {Hirohata}}, \bibinfo {author}
  {\bibfnamefont {S.}~\bibnamefont {Mangin}}, \bibinfo {author} {\bibfnamefont
  {S.~O.}\ \bibnamefont {Valenzuela}}, \bibinfo {author} {\bibfnamefont
  {M.~C.}\ \bibnamefont {Onbaşlı}}, \bibinfo {author} {\bibfnamefont
  {M.}~\bibnamefont {d’Aquino}}, \bibinfo {author} {\bibfnamefont
  {G.}~\bibnamefont {Prenat}}, \bibinfo {author} {\bibfnamefont
  {G.}~\bibnamefont {Finocchio}}, \bibinfo {author} {\bibfnamefont
  {L.}~\bibnamefont {Lopez-Diaz}}, \bibinfo {author} {\bibfnamefont
  {R.}~\bibnamefont {Chantrell}}, \bibinfo {author} {\bibfnamefont
  {O.}~\bibnamefont {Chubykalo-Fesenko}},\ and\ \bibinfo {author}
  {\bibfnamefont {P.}~\bibnamefont {Bortolotti}},\ }\bibfield  {title}
  {\bibinfo {title} {Opportunities and challenges for spintronics in the
  microelectronics industry},\ }\href
  {https://doi.org/10.1038/s41928-020-0461-5} {\bibfield  {journal} {\bibinfo
  {journal} {Nature Electronics}\ }\textbf {\bibinfo {volume} {3}},\ \bibinfo
  {pages} {446} (\bibinfo {year} {2020})}\BibitemShut {NoStop}%
\bibitem [{\citenamefont {Bhatti}\ \emph {et~al.}(2017)\citenamefont {Bhatti},
  \citenamefont {Sbiaa}, \citenamefont {Hirohata}, \citenamefont {Ohno},
  \citenamefont {Fukami},\ and\ \citenamefont
  {Piramanayagam}}]{bhatti_spintronics_2017}%
  \BibitemOpen
  \bibfield  {author} {\bibinfo {author} {\bibfnamefont {S.}~\bibnamefont
  {Bhatti}}, \bibinfo {author} {\bibfnamefont {R.}~\bibnamefont {Sbiaa}},
  \bibinfo {author} {\bibfnamefont {A.}~\bibnamefont {Hirohata}}, \bibinfo
  {author} {\bibfnamefont {H.}~\bibnamefont {Ohno}}, \bibinfo {author}
  {\bibfnamefont {S.}~\bibnamefont {Fukami}},\ and\ \bibinfo {author}
  {\bibfnamefont {S.}~\bibnamefont {Piramanayagam}},\ }\bibfield  {title}
  {\bibinfo {title} {Spintronics based random access memory: a review},\ }\href
  {https://doi.org/10.1016/j.mattod.2017.07.007} {\bibfield  {journal}
  {\bibinfo  {journal} {Materials Today}\ }\textbf {\bibinfo {volume} {20}},\
  \bibinfo {pages} {530} (\bibinfo {year} {2017})}\BibitemShut {NoStop}%
\bibitem [{\citenamefont {Duine}\ \emph {et~al.}(2018)\citenamefont {Duine},
  \citenamefont {Lee}, \citenamefont {Parkin},\ and\ \citenamefont
  {Stiles}}]{duine_synthetic_2018}%
  \BibitemOpen
  \bibfield  {author} {\bibinfo {author} {\bibfnamefont {R.~A.}\ \bibnamefont
  {Duine}}, \bibinfo {author} {\bibfnamefont {K.-J.}\ \bibnamefont {Lee}},
  \bibinfo {author} {\bibfnamefont {S.~S.~P.}\ \bibnamefont {Parkin}},\ and\
  \bibinfo {author} {\bibfnamefont {M.~D.}\ \bibnamefont {Stiles}},\ }\bibfield
   {title} {\bibinfo {title} {Synthetic antiferromagnetic spintronics},\ }\href
  {https://doi.org/10.1038/s41567-018-0050-y} {\bibfield  {journal} {\bibinfo
  {journal} {Nature Physics}\ }\textbf {\bibinfo {volume} {14}},\ \bibinfo
  {pages} {217} (\bibinfo {year} {2018})}\BibitemShut {NoStop}%
\bibitem [{\citenamefont {Chavent}\ \emph {et~al.}(2020)\citenamefont
  {Chavent}, \citenamefont {Iurchuk}, \citenamefont {Tillie}, \citenamefont
  {Bel}, \citenamefont {Lamard}, \citenamefont {Vila}, \citenamefont {Ebels},
  \citenamefont {Sousa}, \citenamefont {Dieny}, \citenamefont {di~Pendina},
  \citenamefont {Prenat}, \citenamefont {Langer}, \citenamefont {Wrona},\ and\
  \citenamefont {Prejbeanu}}]{chavent_multifunctional_2020}%
  \BibitemOpen
  \bibfield  {author} {\bibinfo {author} {\bibfnamefont {A.}~\bibnamefont
  {Chavent}}, \bibinfo {author} {\bibfnamefont {V.}~\bibnamefont {Iurchuk}},
  \bibinfo {author} {\bibfnamefont {L.}~\bibnamefont {Tillie}}, \bibinfo
  {author} {\bibfnamefont {Y.}~\bibnamefont {Bel}}, \bibinfo {author}
  {\bibfnamefont {N.}~\bibnamefont {Lamard}}, \bibinfo {author} {\bibfnamefont
  {L.}~\bibnamefont {Vila}}, \bibinfo {author} {\bibfnamefont {U.}~\bibnamefont
  {Ebels}}, \bibinfo {author} {\bibfnamefont {R.~C.}\ \bibnamefont {Sousa}},
  \bibinfo {author} {\bibfnamefont {B.}~\bibnamefont {Dieny}}, \bibinfo
  {author} {\bibfnamefont {G.}~\bibnamefont {di~Pendina}}, \bibinfo {author}
  {\bibfnamefont {G.}~\bibnamefont {Prenat}}, \bibinfo {author} {\bibfnamefont
  {J.}~\bibnamefont {Langer}}, \bibinfo {author} {\bibfnamefont
  {J.}~\bibnamefont {Wrona}},\ and\ \bibinfo {author} {\bibfnamefont {I.~L.}\
  \bibnamefont {Prejbeanu}},\ }\bibfield  {title} {\bibinfo {title} {A
  multifunctional standardized magnetic tunnel junction stack embedding sensor,
  memory and oscillator functionality},\ }\href
  {https://doi.org/10.1016/j.jmmm.2020.166647} {\bibfield  {journal} {\bibinfo
  {journal} {Journal of Magnetism and Magnetic Materials}\ ,\ \bibinfo {pages}
  {166647}} (\bibinfo {year} {2020})}\BibitemShut {NoStop}%
\bibitem [{\citenamefont {Sousa}\ \emph {et~al.}(2020)\citenamefont {Sousa},
  \citenamefont {Chavent}, \citenamefont {Iurchuk}, \citenamefont {Vila},
  \citenamefont {Ebels}, \citenamefont {Dieny}, \citenamefont {di~Pendina},
  \citenamefont {Prenat}, \citenamefont {Langer}, \citenamefont {Wrona},\ and\
  \citenamefont {Prejbeanu}}]{sousa_2020}%
  \BibitemOpen
  \bibfield  {author} {\bibinfo {author} {\bibfnamefont {R.~C.}\ \bibnamefont
  {Sousa}}, \bibinfo {author} {\bibfnamefont {A.}~\bibnamefont {Chavent}},
  \bibinfo {author} {\bibfnamefont {V.}~\bibnamefont {Iurchuk}}, \bibinfo
  {author} {\bibfnamefont {L.}~\bibnamefont {Vila}}, \bibinfo {author}
  {\bibfnamefont {U.}~\bibnamefont {Ebels}}, \bibinfo {author} {\bibfnamefont
  {B.}~\bibnamefont {Dieny}}, \bibinfo {author} {\bibfnamefont
  {G.}~\bibnamefont {di~Pendina}}, \bibinfo {author} {\bibfnamefont
  {G.}~\bibnamefont {Prenat}}, \bibinfo {author} {\bibfnamefont
  {J.}~\bibnamefont {Langer}}, \bibinfo {author} {\bibfnamefont
  {J.}~\bibnamefont {Wrona}},\ and\ \bibinfo {author} {\bibfnamefont {I.~L.}\
  \bibnamefont {Prejbeanu}},\ }\bibfield  {title} {\bibinfo {title} {Magnetic
  random access memories ({MRAM}) beyond information storage},\ }in\ \href
  {https://doi.org/10.1109/VLSITechnology18217.2020.9265053} {\emph {\bibinfo
  {booktitle} {2020 IEEE Symposium on VLSI Technology}}}\ (\bibinfo {year}
  {2020})\ pp.\ \bibinfo {pages} {1--2}\BibitemShut {NoStop}%
\bibitem [{\citenamefont {Ma}\ \emph {et~al.}(2021)\citenamefont {Ma},
  \citenamefont {Sidi El~Valli}, \citenamefont {Kreißig}, \citenamefont
  {Di~Pendina}, \citenamefont {Protze}, \citenamefont {Ebels}, \citenamefont
  {Prenat}, \citenamefont {Chavent}, \citenamefont {Iurchuk}, \citenamefont
  {Sousa}, \citenamefont {Vila}, \citenamefont {Ellinger}, \citenamefont
  {Langer}, \citenamefont {Wrona},\ and\ \citenamefont {Prejbeanu}}]{ma_2021}%
  \BibitemOpen
  \bibfield  {author} {\bibinfo {author} {\bibfnamefont {R.}~\bibnamefont
  {Ma}}, \bibinfo {author} {\bibfnamefont {A.}~\bibnamefont {Sidi El~Valli}},
  \bibinfo {author} {\bibfnamefont {M.}~\bibnamefont {Kreißig}}, \bibinfo
  {author} {\bibfnamefont {G.}~\bibnamefont {Di~Pendina}}, \bibinfo {author}
  {\bibfnamefont {F.}~\bibnamefont {Protze}}, \bibinfo {author} {\bibfnamefont
  {U.}~\bibnamefont {Ebels}}, \bibinfo {author} {\bibfnamefont
  {G.}~\bibnamefont {Prenat}}, \bibinfo {author} {\bibfnamefont
  {A.}~\bibnamefont {Chavent}}, \bibinfo {author} {\bibfnamefont
  {V.}~\bibnamefont {Iurchuk}}, \bibinfo {author} {\bibfnamefont
  {R.}~\bibnamefont {Sousa}}, \bibinfo {author} {\bibfnamefont
  {L.}~\bibnamefont {Vila}}, \bibinfo {author} {\bibfnamefont {F.}~\bibnamefont
  {Ellinger}}, \bibinfo {author} {\bibfnamefont {J.}~\bibnamefont {Langer}},
  \bibinfo {author} {\bibfnamefont {J.}~\bibnamefont {Wrona}},\ and\ \bibinfo
  {author} {\bibfnamefont {I.-L.}\ \bibnamefont {Prejbeanu}},\ }\bibfield
  {title} {\bibinfo {title} {Microwave functionality of spintronic devices
  implemented in a hybrid complementary metal oxide semiconductor and magnetic
  tunnel junction technology},\ }\href
  {https://doi.org/https://doi.org/10.1049/ell2.12103} {\bibfield  {journal}
  {\bibinfo  {journal} {Electronics Letters}\ }\textbf {\bibinfo {volume}
  {57}},\ \bibinfo {pages} {264} (\bibinfo {year} {2021})}\BibitemShut
  {NoStop}%
\bibitem [{\citenamefont {Binasch}\ \emph {et~al.}(1989)\citenamefont
  {Binasch}, \citenamefont {Grünberg}, \citenamefont {Saurenbach},\ and\
  \citenamefont {Zinn}}]{Grunberg_GMR_1989}%
  \BibitemOpen
  \bibfield  {author} {\bibinfo {author} {\bibfnamefont {G.}~\bibnamefont
  {Binasch}}, \bibinfo {author} {\bibfnamefont {P.}~\bibnamefont {Grünberg}},
  \bibinfo {author} {\bibfnamefont {F.}~\bibnamefont {Saurenbach}},\ and\
  \bibinfo {author} {\bibfnamefont {W.}~\bibnamefont {Zinn}},\ }\bibfield
  {title} {\bibinfo {title} {Enhanced magnetoresistance in layered magnetic
  structures with antiferromagnetic interlayer exchange},\ }\href
  {https://doi.org/10.1103/PhysRevB.39.4828} {\bibfield  {journal} {\bibinfo
  {journal} {Physical Review B}\ }\textbf {\bibinfo {volume} {39}},\ \bibinfo
  {pages} {4828} (\bibinfo {year} {1989})}\BibitemShut {NoStop}%
\bibitem [{\citenamefont {Baibich}\ \emph {et~al.}(1988)\citenamefont
  {Baibich}, \citenamefont {Broto}, \citenamefont {Fert}, \citenamefont
  {Van~Dau}, \citenamefont {Petroff}, \citenamefont {Etienne}, \citenamefont
  {Creuzet}, \citenamefont {Friederich},\ and\ \citenamefont
  {Chazelas}}]{Fert_GMR_1988}%
  \BibitemOpen
  \bibfield  {author} {\bibinfo {author} {\bibfnamefont {M.~N.}\ \bibnamefont
  {Baibich}}, \bibinfo {author} {\bibfnamefont {J.~M.}\ \bibnamefont {Broto}},
  \bibinfo {author} {\bibfnamefont {A.}~\bibnamefont {Fert}}, \bibinfo {author}
  {\bibfnamefont {F.~N.}\ \bibnamefont {Van~Dau}}, \bibinfo {author}
  {\bibfnamefont {F.}~\bibnamefont {Petroff}}, \bibinfo {author} {\bibfnamefont
  {P.}~\bibnamefont {Etienne}}, \bibinfo {author} {\bibfnamefont
  {G.}~\bibnamefont {Creuzet}}, \bibinfo {author} {\bibfnamefont
  {A.}~\bibnamefont {Friederich}},\ and\ \bibinfo {author} {\bibfnamefont
  {J.}~\bibnamefont {Chazelas}},\ }\bibfield  {title} {\bibinfo {title} {Giant
  {Magnetoresistance} of (001){Fe}/(001){Cr} {Magnetic} {Superlattices}},\
  }\href {https://doi.org/10.1103/PhysRevLett.61.2472} {\bibfield  {journal}
  {\bibinfo  {journal} {Physical Review Letters}\ }\textbf {\bibinfo {volume}
  {61}},\ \bibinfo {pages} {2472} (\bibinfo {year} {1988})}\BibitemShut
  {NoStop}%
\bibitem [{\citenamefont {Stiles}(2004)}]{Stiles_interlayer_2004}%
  \BibitemOpen
  \bibfield  {author} {\bibinfo {author} {\bibfnamefont {M.}~\bibnamefont
  {Stiles}},\ }\bibinfo {title} {Interlayer exchange coupling}\ (\bibinfo
  {publisher} {Ultrathin Magnetic Structures III, Springer-Verlag, New York,
  NY},\ \bibinfo {year} {2004})\ Chap.~\bibinfo {chapter} {4}, pp.\ \bibinfo
  {pages} {99--142}\BibitemShut {NoStop}%
\bibitem [{\citenamefont {Ruderman}\ and\ \citenamefont
  {Kittel}(1954)}]{ruderman_kittel_RKKY_1954}%
  \BibitemOpen
  \bibfield  {author} {\bibinfo {author} {\bibfnamefont {M.~A.}\ \bibnamefont
  {Ruderman}}\ and\ \bibinfo {author} {\bibfnamefont {C.}~\bibnamefont
  {Kittel}},\ }\bibfield  {title} {\bibinfo {title} {Indirect {Exchange}
  {Coupling} of {Nuclear} {Magnetic} {Moments} by {Conduction} {Electrons}},\
  }\href {https://doi.org/10.1103/PhysRev.96.99} {\bibfield  {journal}
  {\bibinfo  {journal} {Physical Review}\ }\textbf {\bibinfo {volume} {96}},\
  \bibinfo {pages} {99} (\bibinfo {year} {1954})}\BibitemShut {NoStop}%
\bibitem [{\citenamefont {Kasuya}(1956)}]{kasuya_RKKY_1956}%
  \BibitemOpen
  \bibfield  {author} {\bibinfo {author} {\bibfnamefont {T.}~\bibnamefont
  {Kasuya}},\ }\bibfield  {title} {\bibinfo {title} {A {Theory} of {Metallic}
  {Ferro}- and {Antiferromagnetism} on {Zener}'s {Model}},\ }\href
  {https://doi.org/10.1143/PTP.16.45} {\bibfield  {journal} {\bibinfo
  {journal} {Progress of Theoretical Physics}\ }\textbf {\bibinfo {volume}
  {16}},\ \bibinfo {pages} {45} (\bibinfo {year} {1956})}\BibitemShut {NoStop}%
\bibitem [{\citenamefont {Yosida}(1957)}]{yosida_RKKY_1957}%
  \BibitemOpen
  \bibfield  {author} {\bibinfo {author} {\bibfnamefont {K.}~\bibnamefont
  {Yosida}},\ }\bibfield  {title} {\bibinfo {title} {Magnetic {Properties} of
  {Cu}-{Mn} {Alloys}},\ }\href {https://doi.org/10.1103/PhysRev.106.893}
  {\bibfield  {journal} {\bibinfo  {journal} {Physical Review}\ }\textbf
  {\bibinfo {volume} {106}},\ \bibinfo {pages} {893} (\bibinfo {year}
  {1957})}\BibitemShut {NoStop}%
\bibitem [{\citenamefont {Bruno}(1995)}]{bruno_theory_1995}%
  \BibitemOpen
  \bibfield  {author} {\bibinfo {author} {\bibfnamefont {P.}~\bibnamefont
  {Bruno}},\ }\bibfield  {title} {\bibinfo {title} {Theory of interlayer
  magnetic coupling},\ }\href {https://doi.org/10.1103/PhysRevB.52.411}
  {\bibfield  {journal} {\bibinfo  {journal} {Phys. Rev. B}\ }\textbf {\bibinfo
  {volume} {52}},\ \bibinfo {pages} {411} (\bibinfo {year} {1995})}\BibitemShut
  {NoStop}%
\bibitem [{\citenamefont {Babic}\ \emph {et~al.}(1980)\citenamefont {Babic},
  \citenamefont {Kajzar},\ and\ \citenamefont
  {Parette}}]{babicIronMomentChromiumrich1980}%
  \BibitemOpen
  \bibfield  {author} {\bibinfo {author} {\bibfnamefont {B.}~\bibnamefont
  {Babic}}, \bibinfo {author} {\bibfnamefont {F.}~\bibnamefont {Kajzar}},\ and\
  \bibinfo {author} {\bibfnamefont {G.}~\bibnamefont {Parette}},\ }\bibfield
  {title} {\bibinfo {title} {Iron moment in chromium-rich {Cr}-{Fe} alloys},\
  }\href {https://doi.org/10.1016/0304-8853(80)91053-7} {\bibfield  {journal}
  {\bibinfo  {journal} {Journal of Magnetism and Magnetic Materials}\ }\textbf
  {\bibinfo {volume} {15-18}},\ \bibinfo {pages} {287} (\bibinfo {year}
  {1980})}\BibitemShut {NoStop}%
\bibitem [{\citenamefont {Burke}\ \emph {et~al.}(1978)\citenamefont {Burke},
  \citenamefont {Cywinski},\ and\ \citenamefont
  {Rainford}}]{burkeSuperparamagnetismCharacterMagnetic1978}%
  \BibitemOpen
  \bibfield  {author} {\bibinfo {author} {\bibfnamefont {S.~K.}\ \bibnamefont
  {Burke}}, \bibinfo {author} {\bibfnamefont {R.}~\bibnamefont {Cywinski}},\
  and\ \bibinfo {author} {\bibfnamefont {B.~D.}\ \bibnamefont {Rainford}},\
  }\bibfield  {title} {\bibinfo {title} {Superparamagnetism and the character
  of magnetic order in binary {Cr}–{Fe} alloys near the critical
  concentration},\ }\href {https://doi.org/10.1107/S0021889878014120}
  {\bibfield  {journal} {\bibinfo  {journal} {Journal of Applied
  Crystallography}\ }\textbf {\bibinfo {volume} {11}},\ \bibinfo {pages} {644}
  (\bibinfo {year} {1978})}\BibitemShut {NoStop}%
\bibitem [{\citenamefont {Polishchuk}\ \emph {et~al.}(2017)\citenamefont
  {Polishchuk}, \citenamefont {Tykhonenko-Polishchuk}, \citenamefont
  {Holmgren}, \citenamefont {Kravets},\ and\ \citenamefont
  {Korenivski}}]{polishchuk_thermally_2017}%
  \BibitemOpen
  \bibfield  {author} {\bibinfo {author} {\bibfnamefont {D.~M.}\ \bibnamefont
  {Polishchuk}}, \bibinfo {author} {\bibfnamefont {Y.~O.}\ \bibnamefont
  {Tykhonenko-Polishchuk}}, \bibinfo {author} {\bibfnamefont {E.}~\bibnamefont
  {Holmgren}}, \bibinfo {author} {\bibfnamefont {A.~F.}\ \bibnamefont
  {Kravets}},\ and\ \bibinfo {author} {\bibfnamefont {V.}~\bibnamefont
  {Korenivski}},\ }\bibfield  {title} {\bibinfo {title} {Thermally induced
  antiferromagnetic exchange in magnetic multilayers},\ }\href
  {https://doi.org/10.1103/PhysRevB.96.104427} {\bibfield  {journal} {\bibinfo
  {journal} {Physical Review B}\ }\textbf {\bibinfo {volume} {96}},\ \bibinfo
  {pages} {104427} (\bibinfo {year} {2017})}\BibitemShut {NoStop}%
\bibitem [{\citenamefont {Polishchuk}\ \emph
  {et~al.}(2018{\natexlab{a}})\citenamefont {Polishchuk}, \citenamefont
  {Tykhonenko-Polishchuk}, \citenamefont {Holmgren}, \citenamefont {Kravets},
  \citenamefont {Tovstolytkin},\ and\ \citenamefont
  {Korenivski}}]{polishchuk_giant_2018}%
  \BibitemOpen
  \bibfield  {author} {\bibinfo {author} {\bibfnamefont {D.~M.}\ \bibnamefont
  {Polishchuk}}, \bibinfo {author} {\bibfnamefont {Y.~O.}\ \bibnamefont
  {Tykhonenko-Polishchuk}}, \bibinfo {author} {\bibfnamefont {E.}~\bibnamefont
  {Holmgren}}, \bibinfo {author} {\bibfnamefont {A.~F.}\ \bibnamefont
  {Kravets}}, \bibinfo {author} {\bibfnamefont {A.~I.}\ \bibnamefont
  {Tovstolytkin}},\ and\ \bibinfo {author} {\bibfnamefont {V.}~\bibnamefont
  {Korenivski}},\ }\bibfield  {title} {\bibinfo {title} {Giant magnetocaloric
  effect driven by indirect exchange in magnetic multilayers},\ }\href
  {https://doi.org/10.1103/PhysRevMaterials.2.114402} {\bibfield  {journal}
  {\bibinfo  {journal} {Physical Review Materials}\ }\textbf {\bibinfo {volume}
  {2}},\ \bibinfo {pages} {114402} (\bibinfo {year}
  {2018}{\natexlab{a}})}\BibitemShut {NoStop}%
\bibitem [{\citenamefont {Polishchuk}\ \emph {et~al.}(2021)\citenamefont
  {Polishchuk}, \citenamefont {Tykhonenko-Polishchuk}, \citenamefont
  {Lytvynenko}, \citenamefont {Rostas}, \citenamefont {Gomonay},\ and\
  \citenamefont {Korenivski}}]{polishchuk_thermal_2021}%
  \BibitemOpen
  \bibfield  {author} {\bibinfo {author} {\bibfnamefont {D.}~\bibnamefont
  {Polishchuk}}, \bibinfo {author} {\bibfnamefont {Y.}~\bibnamefont
  {Tykhonenko-Polishchuk}}, \bibinfo {author} {\bibfnamefont {Y.}~\bibnamefont
  {Lytvynenko}}, \bibinfo {author} {\bibfnamefont {A.}~\bibnamefont {Rostas}},
  \bibinfo {author} {\bibfnamefont {O.}~\bibnamefont {Gomonay}},\ and\ \bibinfo
  {author} {\bibfnamefont {V.}~\bibnamefont {Korenivski}},\ }\bibfield  {title}
  {\bibinfo {title} {Thermal gating of magnon exchange in magnetic multilayers
  with antiferromagnetic spacers},\ }\href
  {https://doi.org/10.1103/PhysRevLett.126.227203} {\bibfield  {journal}
  {\bibinfo  {journal} {Physical Review Letters}\ }\textbf {\bibinfo {volume}
  {126}},\ \bibinfo {pages} {227203} (\bibinfo {year} {2021})}\BibitemShut
  {NoStop}%
\bibitem [{\citenamefont {Valet}\ and\ \citenamefont
  {Fert}(1993)}]{valet_1993}%
  \BibitemOpen
  \bibfield  {author} {\bibinfo {author} {\bibfnamefont {T.}~\bibnamefont
  {Valet}}\ and\ \bibinfo {author} {\bibfnamefont {A.}~\bibnamefont {Fert}},\
  }\bibfield  {title} {\bibinfo {title} {Theory of the perpendicular
  magnetoresistance in magnetic multilayers},\ }\href
  {https://doi.org/10.1103/PhysRevB.48.7099} {\bibfield  {journal} {\bibinfo
  {journal} {Phys. Rev. B}\ }\textbf {\bibinfo {volume} {48}},\ \bibinfo
  {pages} {7099} (\bibinfo {year} {1993})}\BibitemShut {NoStop}%
\bibitem [{\citenamefont {Polishchuk}\ \emph
  {et~al.}(2018{\natexlab{b}})\citenamefont {Polishchuk}, \citenamefont
  {Tykhonenko-Polishchuk}, \citenamefont {Borynskyi}, \citenamefont {Kravets},
  \citenamefont {Tovstolytkin},\ and\ \citenamefont
  {Korenivski}}]{polishchukMagneticHysteresisNanostructures2018}%
  \BibitemOpen
  \bibfield  {author} {\bibinfo {author} {\bibfnamefont {D.}~\bibnamefont
  {Polishchuk}}, \bibinfo {author} {\bibfnamefont {Y.}~\bibnamefont
  {Tykhonenko-Polishchuk}}, \bibinfo {author} {\bibfnamefont {V.}~\bibnamefont
  {Borynskyi}}, \bibinfo {author} {\bibfnamefont {A.}~\bibnamefont {Kravets}},
  \bibinfo {author} {\bibfnamefont {A.}~\bibnamefont {Tovstolytkin}},\ and\
  \bibinfo {author} {\bibfnamefont {V.}~\bibnamefont {Korenivski}},\ }\bibfield
   {title} {\bibinfo {title} {Magnetic {Hysteresis} in {Nanostructures} with
  {Thermally} {Controlled} {RKKY} {Coupling}},\ }\href
  {https://doi.org/10.1186/s11671-018-2669-0} {\bibfield  {journal} {\bibinfo
  {journal} {Nanoscale Research Letters}\ }\textbf {\bibinfo {volume} {13}},\
  \bibinfo {pages} {245} (\bibinfo {year} {2018}{\natexlab{b}})}\BibitemShut
  {NoStop}%
\bibitem [{\citenamefont {Belmeguenai}\ \emph {et~al.}(2007)\citenamefont
  {Belmeguenai}, \citenamefont {Martin}, \citenamefont {Woltersdorf},
  \citenamefont {Maier},\ and\ \citenamefont
  {Bayreuther}}]{Belmeguenai_model_2007}%
  \BibitemOpen
  \bibfield  {author} {\bibinfo {author} {\bibfnamefont {M.}~\bibnamefont
  {Belmeguenai}}, \bibinfo {author} {\bibfnamefont {T.}~\bibnamefont {Martin}},
  \bibinfo {author} {\bibfnamefont {G.}~\bibnamefont {Woltersdorf}}, \bibinfo
  {author} {\bibfnamefont {M.}~\bibnamefont {Maier}},\ and\ \bibinfo {author}
  {\bibfnamefont {G.}~\bibnamefont {Bayreuther}},\ }\bibfield  {title}
  {\bibinfo {title} {Frequency- and time-domain investigation of the dynamic
  properties of interlayer-exchange-coupled
  {Ni}$_{81}${Fe}$_{19}$/{Ru}/{Ni}$_{81}${Fe}$_{19}$ thin films},\ }\href
  {https://doi.org/10.1103/PhysRevB.76.104414} {\bibfield  {journal} {\bibinfo
  {journal} {Phys. Rev. B}\ }\textbf {\bibinfo {volume} {76}},\ \bibinfo
  {pages} {104414} (\bibinfo {year} {2007})}\BibitemShut {NoStop}%
\bibitem [{\citenamefont {Sorokin}\ \emph {et~al.}(2020)\citenamefont
  {Sorokin}, \citenamefont {Gallardo}, \citenamefont {Fowley}, \citenamefont
  {Lenz}, \citenamefont {Titova}, \citenamefont {Atcheson}, \citenamefont
  {Dennehy}, \citenamefont {Rode}, \citenamefont {Fassbender}, \citenamefont
  {Lindner},\ and\ \citenamefont {Deac}}]{sorokin_magnetization_2020}%
  \BibitemOpen
  \bibfield  {author} {\bibinfo {author} {\bibfnamefont {S.}~\bibnamefont
  {Sorokin}}, \bibinfo {author} {\bibfnamefont {R.~A.}\ \bibnamefont
  {Gallardo}}, \bibinfo {author} {\bibfnamefont {C.}~\bibnamefont {Fowley}},
  \bibinfo {author} {\bibfnamefont {K.}~\bibnamefont {Lenz}}, \bibinfo {author}
  {\bibfnamefont {A.}~\bibnamefont {Titova}}, \bibinfo {author} {\bibfnamefont
  {G.~Y.~P.}\ \bibnamefont {Atcheson}}, \bibinfo {author} {\bibfnamefont
  {G.}~\bibnamefont {Dennehy}}, \bibinfo {author} {\bibfnamefont
  {K.}~\bibnamefont {Rode}}, \bibinfo {author} {\bibfnamefont {J.}~\bibnamefont
  {Fassbender}}, \bibinfo {author} {\bibfnamefont {J.}~\bibnamefont
  {Lindner}},\ and\ \bibinfo {author} {\bibfnamefont {A.~M.}\ \bibnamefont
  {Deac}},\ }\bibfield  {title} {\bibinfo {title} {Magnetization dynamics in
  synthetic antiferromagnets: Role of dynamical energy and mutual spin
  pumping},\ }\href {https://doi.org/10.1103/PhysRevB.101.144410} {\bibfield
  {journal} {\bibinfo  {journal} {Physical Review B}\ }\textbf {\bibinfo
  {volume} {101}},\ \bibinfo {pages} {144410} (\bibinfo {year}
  {2020})}\BibitemShut {NoStop}%
\bibitem [{\citenamefont
  {Slonczewski}(1993)}]{slonczewskiOriginBiquadraticExchange1993}%
  \BibitemOpen
  \bibfield  {author} {\bibinfo {author} {\bibfnamefont {J.~C.}\ \bibnamefont
  {Slonczewski}},\ }\bibfield  {title} {\bibinfo {title} {Origin of biquadratic
  exchange in magnetic multilayers (invited)},\ }\href
  {https://doi.org/10.1063/1.353483} {\bibfield  {journal} {\bibinfo  {journal}
  {Journal of Applied Physics}\ }\textbf {\bibinfo {volume} {73}},\ \bibinfo
  {pages} {5957} (\bibinfo {year} {1993})}\BibitemShut {NoStop}%
\bibitem [{\citenamefont {Prejbeanu}\ \emph {et~al.}(2007)\citenamefont
  {Prejbeanu}, \citenamefont {Kerekes}, \citenamefont {Sousa}, \citenamefont
  {Sibuet}, \citenamefont {Redon}, \citenamefont {Dieny},\ and\ \citenamefont
  {Nozières}}]{prejbeanuThermallyAssistedMRAM2007}%
  \BibitemOpen
  \bibfield  {author} {\bibinfo {author} {\bibfnamefont {I.~L.}\ \bibnamefont
  {Prejbeanu}}, \bibinfo {author} {\bibfnamefont {M.}~\bibnamefont {Kerekes}},
  \bibinfo {author} {\bibfnamefont {R.~C.}\ \bibnamefont {Sousa}}, \bibinfo
  {author} {\bibfnamefont {H.}~\bibnamefont {Sibuet}}, \bibinfo {author}
  {\bibfnamefont {O.}~\bibnamefont {Redon}}, \bibinfo {author} {\bibfnamefont
  {B.}~\bibnamefont {Dieny}},\ and\ \bibinfo {author} {\bibfnamefont {J.~P.}\
  \bibnamefont {Nozières}},\ }\bibfield  {title} {\bibinfo {title} {Thermally
  assisted {MRAM}},\ }\href {https://doi.org/10.1088/0953-8984/19/16/165218}
  {\bibfield  {journal} {\bibinfo  {journal} {Journal of Physics: Condensed
  Matter}\ }\textbf {\bibinfo {volume} {19}},\ \bibinfo {pages} {165218}
  (\bibinfo {year} {2007})}\BibitemShut {NoStop}%
\bibitem [{\citenamefont {Le~Gallo}\ \emph {et~al.}(2016)\citenamefont
  {Le~Gallo}, \citenamefont {Athmanathan}, \citenamefont {Krebs},\ and\
  \citenamefont {Sebastian}}]{legalloEvidenceThermallyAssisted2016}%
  \BibitemOpen
  \bibfield  {author} {\bibinfo {author} {\bibfnamefont {M.}~\bibnamefont
  {Le~Gallo}}, \bibinfo {author} {\bibfnamefont {A.}~\bibnamefont
  {Athmanathan}}, \bibinfo {author} {\bibfnamefont {D.}~\bibnamefont {Krebs}},\
  and\ \bibinfo {author} {\bibfnamefont {A.}~\bibnamefont {Sebastian}},\
  }\bibfield  {title} {\bibinfo {title} {Evidence for thermally assisted
  threshold switching behavior in nanoscale phase-change memory cells},\ }\href
  {https://doi.org/10.1063/1.4938532} {\bibfield  {journal} {\bibinfo
  {journal} {Journal of Applied Physics}\ }\textbf {\bibinfo {volume} {119}},\
  \bibinfo {pages} {025704} (\bibinfo {year} {2016})}\BibitemShut {NoStop}%
\bibitem [{\citenamefont {Strelkov}\ \emph {et~al.}(2018)\citenamefont
  {Strelkov}, \citenamefont {Chavent}, \citenamefont {Timopheev}, \citenamefont
  {Sousa}, \citenamefont {Prejbeanu}, \citenamefont {Buda-Prejbeanu},\ and\
  \citenamefont {Dieny}}]{strelkov_2018}%
  \BibitemOpen
  \bibfield  {author} {\bibinfo {author} {\bibfnamefont {N.}~\bibnamefont
  {Strelkov}}, \bibinfo {author} {\bibfnamefont {A.}~\bibnamefont {Chavent}},
  \bibinfo {author} {\bibfnamefont {A.}~\bibnamefont {Timopheev}}, \bibinfo
  {author} {\bibfnamefont {R.~C.}\ \bibnamefont {Sousa}}, \bibinfo {author}
  {\bibfnamefont {I.~L.}\ \bibnamefont {Prejbeanu}}, \bibinfo {author}
  {\bibfnamefont {L.~D.}\ \bibnamefont {Buda-Prejbeanu}},\ and\ \bibinfo
  {author} {\bibfnamefont {B.}~\bibnamefont {Dieny}},\ }\bibfield  {title}
  {\bibinfo {title} {Impact of joule heating on the stability phase diagrams of
  perpendicular magnetic tunnel junctions},\ }\href
  {https://doi.org/10.1103/PhysRevB.98.214410} {\bibfield  {journal} {\bibinfo
  {journal} {Phys. Rev. B}\ }\textbf {\bibinfo {volume} {98}},\ \bibinfo
  {pages} {214410} (\bibinfo {year} {2018})}\BibitemShut {NoStop}%
\bibitem [{\citenamefont {Chavent}\ \emph {et~al.}(2016)\citenamefont
  {Chavent}, \citenamefont {Ducruet}, \citenamefont {Portemont}, \citenamefont
  {Vila}, \citenamefont {Alvarez-Hérault}, \citenamefont {Sousa},
  \citenamefont {Prejbeanu},\ and\ \citenamefont
  {Dieny}}]{chaventSteadyStateDynamics2016}%
  \BibitemOpen
  \bibfield  {author} {\bibinfo {author} {\bibfnamefont {A.}~\bibnamefont
  {Chavent}}, \bibinfo {author} {\bibfnamefont {C.}~\bibnamefont {Ducruet}},
  \bibinfo {author} {\bibfnamefont {C.}~\bibnamefont {Portemont}}, \bibinfo
  {author} {\bibfnamefont {L.}~\bibnamefont {Vila}}, \bibinfo {author}
  {\bibfnamefont {J.}~\bibnamefont {Alvarez-Hérault}}, \bibinfo {author}
  {\bibfnamefont {R.}~\bibnamefont {Sousa}}, \bibinfo {author} {\bibfnamefont
  {I.}~\bibnamefont {Prejbeanu}},\ and\ \bibinfo {author} {\bibfnamefont
  {B.}~\bibnamefont {Dieny}},\ }\bibfield  {title} {\bibinfo {title} {Steady
  {State} and {Dynamics} of {Joule} {Heating} in {Magnetic} {Tunnel}
  {Junctions} {Observed} via the {Temperature} {Dependence} of {RKKY}
  {Coupling}},\ }\href {https://doi.org/10.1103/PhysRevApplied.6.034003}
  {\bibfield  {journal} {\bibinfo  {journal} {Physical Review Applied}\
  }\textbf {\bibinfo {volume} {6}},\ \bibinfo {pages} {034003} (\bibinfo {year}
  {2016})}\BibitemShut {NoStop}%
\bibitem [{\citenamefont {Wu}\ \emph {et~al.}(2022)\citenamefont {Wu},
  \citenamefont {Zhang}, \citenamefont {Wang}, \citenamefont {Groß},
  \citenamefont {Yang}, \citenamefont {Li}, \citenamefont {Guo}, \citenamefont
  {He}, \citenamefont {Wong}, \citenamefont {Wu}, \citenamefont {Han},
  \citenamefont {Lai}, \citenamefont {Gräfe}, \citenamefont {Cheng},\ and\
  \citenamefont {Wang}}]{wuCurrentinducedNeelOrder2022}%
  \BibitemOpen
  \bibfield  {author} {\bibinfo {author} {\bibfnamefont {H.}~\bibnamefont
  {Wu}}, \bibinfo {author} {\bibfnamefont {H.}~\bibnamefont {Zhang}}, \bibinfo
  {author} {\bibfnamefont {B.}~\bibnamefont {Wang}}, \bibinfo {author}
  {\bibfnamefont {F.}~\bibnamefont {Groß}}, \bibinfo {author} {\bibfnamefont
  {C.-Y.}\ \bibnamefont {Yang}}, \bibinfo {author} {\bibfnamefont
  {G.}~\bibnamefont {Li}}, \bibinfo {author} {\bibfnamefont {C.}~\bibnamefont
  {Guo}}, \bibinfo {author} {\bibfnamefont {H.}~\bibnamefont {He}}, \bibinfo
  {author} {\bibfnamefont {K.}~\bibnamefont {Wong}}, \bibinfo {author}
  {\bibfnamefont {D.}~\bibnamefont {Wu}}, \bibinfo {author} {\bibfnamefont
  {X.}~\bibnamefont {Han}}, \bibinfo {author} {\bibfnamefont {C.-H.}\
  \bibnamefont {Lai}}, \bibinfo {author} {\bibfnamefont {J.}~\bibnamefont
  {Gräfe}}, \bibinfo {author} {\bibfnamefont {R.}~\bibnamefont {Cheng}},\ and\
  \bibinfo {author} {\bibfnamefont {K.~L.}\ \bibnamefont {Wang}},\ }\bibfield
  {title} {\bibinfo {title} {Current-induced {Néel} order switching
  facilitated by magnetic phase transition},\ }\href
  {https://doi.org/10.1038/s41467-022-29170-2} {\bibfield  {journal} {\bibinfo
  {journal} {Nature Communications}\ }\textbf {\bibinfo {volume} {13}},\
  \bibinfo {pages} {1629} (\bibinfo {year} {2022})}\BibitemShut {NoStop}%
\end{thebibliography}%

\end{document}